\documentclass[aps,prb,twocolumn,showpacs]{revtex4}

\usepackage{graphicx}

\begin{document}

\title{Electronic specific heat and low energy quasiparticle excitations in superconducting state of $La_{2-x}Sr_xCuO_4$ single crystals}

\author{Hai-Hu Wen\email, Zhi-Yong Liu, Fang Zhou, Jiwu Xiong, Wenxing Ti}

\affiliation{National Laboratory for Superconductivity, Institute
of Physics, Chinese Academy of Sciences, P.~O.~Box 603, Beijing
100080, P.~R.~China}
\author{Tao Xiang}
\affiliation{Institute of Theoretical Physics and ICTS, Chinese
Academy of Sciences, P. O. Box 2735, Beijing 100080, P.~R.~China}

\author{Seiki Komiya, Xuefeng Sun, Yoichi Ando}
\affiliation{Central Research Institute of Electric Power
Industry, Komae, Tokyo 201-8511, Japan}

\date{\today}

\begin{abstract}
Low temperature specific heat has been measured and extensively
analyzed on a series of $La_{2-x}Sr_xCuO_4$ single crystals from
underdoped to overdoped regime. From these data the quasiparticle
density of states (DOS) in the mixed state is derived and compared
to the predicted scaling law $C_{vol}/T\sqrt{H}=f(T/\sqrt{H})$ of
d-wave superconductivity. It is found that the scaling law can be
nicely followed by the optimally doped sample (x=0.15) in quite
wide region of ($T/\sqrt{H} \leq 8 K /\sqrt{T}$). However, the
region for this scaling becomes smaller and smaller towards more
underdoped region: a clear trend can be seen for samples from x=
0.15 to 0.069. Therefore, generally speaking, the scaling quality
becomes worse on the underdoped samples in terms of scalable
region of $T/\sqrt{H}$. This feature in the underdoped region is
explained as due to the low energy excitations from a second order
(for example, anti-ferromagnetic correlation, d-density wave, spin
density wave or charge density wave order) that may co-exist or
compete with superconductivity. Surprisingly, deviations from the
d-wave scaling law have also been found for the overdoped sample
(x=0.22). While the scaling law is reconciled for the overdoped
sample when the core size effect is taken into account. An
important discovery of present work is that the zero-temperature
data follow the Volovik's relation $\Delta \gamma(T=0)=A\sqrt{H}$
quite well for all samples investigated here although the
applicability of the d-wave scaling law to the data at finite
temperatures varies with doped hole concentration. Finally we
present the doping dependence of some parameters, such as, the
residual linear term $\gamma_0$, the $\alpha$ value, etc.  It is
suggested that the residual linear term ($\gamma_0T$) of the
electronic specific heat observed in all cuprate superconductors
is probably due to the inhomogeneity, either chemical or
electronic in origin. The field induced reduction of the specific
heat in the mixed state is also reported. Finally implications on
the electronic phase diagram are suggested.

\end{abstract}

\maketitle

\section{Introduction}
One of few points with consensus in the cuprate superconductors is
the $d_{x^2-y^2}$ pairing symmetry in hole doped region. This has
been supported by tremendous experiments \cite{Tsuei1} both from
surface detection\cite{Tsuei2,ARPES,Hardy,Tunneling,YehNC} and
bulk
measurements\cite{NMR,Moler,Revaz,Wright,Fisher,Phillips,Nohara,Chen}.
In a d-wave superconductor with line nodes in the gap function,
the quasiparticle density of states (DOS) $N(E)$ rises linearly
with energy at the Fermi level in zero field, i.e.,
$N(E)\propto|E-E_F|$, resulting in\cite{Kopnin1996} an electronic
specific heat $C_e=\alpha T^2$, where $\alpha \propto
\gamma_n/T_c$ and $\gamma_n$ is the specific heat which is
proportional to the DOS at the Fermi level of the normal state. In
the mixed state with the field higher than a certain value, the
DOS near the Fermi surface becomes finite, therefore the quadratic
term $C_e=\alpha T^2$ will be surpassed and substituted by both
the localized excitations inside the vortex core and the
de-localized excitations outside the core. Volovik \cite{Volovik}
pointed out that for d-wave superconductors in the mixed state,
supercurrents around a vortex core lead to a Doppler shift to the
quasiparticle excitation spectrum, which affects strongly the low
energy excitation around the nodes. It was shown that the
contribution from the delocalized part (outside the core) will
prevail over the core part and the specific heat is predicted to
behave as\cite{Volovik,Kopnin1996}
$C_{vol}=k\gamma_nT\sqrt{H/H_{c2}}$ with $k$ in the order of
unity. This prediction has been verified by many measurements
which were taken as the evidence for d-wave
symmetry\cite{Moler,Revaz,Wright,Fisher,Phillips,Nohara,Chen,Hussey}.
In the finite temperature and field region a scaling law is
proposed\cite{SimonLee} as

\begin{equation}
C_{vol}/T\sqrt{H}=f(T/\sqrt{H})
\end{equation}

with $T/\sqrt{H}\le T_c/\sqrt{H_{c2}(0)}$. This scaling law can be
further converted into the form of $C_{vol}/H=g(T/\sqrt{H})$ or
$C_{vol}/T^2=y(T/\sqrt{H})$, here $f(x)$, or $g(x)$ or $y(x)$ are
unknown scaling functions. This scaling law has been proved in
$YBCO$\cite{Moler,Revaz,Wright} and in
LSCO\cite{Fisher,Phillips,Nohara,Chen}. It remains however unclear
whether this scaling law is still valid in the very overdoped
region since the vortex core size $\xi$ grows up. In the
underdoped region, inelastic neutron scattering reveals that an
anti-ferromagnetic order emerges when the superconductivity is
suppressed\cite{Lake,Kang}. It is thus also interesting to check
whether the d-wave scaling law proposed by Simon and Lee is
applicable in underdoped regime. In addition, the Simon-Lee
scaling law is in agreement with the calculations as proposed by
Volovik and Kopnin\cite{Volovik,Kopnin1996} in two extreme
conditions of temperature. In the low temperature limit the
scaling law $C_{vol}/T^2=y(T/\sqrt{H})$ becomes the Volovik's
relation $C_{vol}=AT\sqrt{H}$. When the temperature is increased,
another relation $C\propto aT^2+bH$ is reached. The boundary
between these two regions is $T/\sqrt{H}=T_c/\sqrt{H_{c2}}$
according to Volovik and Kopnin\cite{VolovikComment}. These
theoretical models can be quantitatively tested by experiments on
samples with different doping concentrations.

Another important but controversial issue is the vortex core state
in the cuprate superconductors. By solving the mean-field
Bogoliubov-de Gennes (BdG) equation, theoretically it is suggested
that a zero-bias conductance peak (ZBCP) exists in the vortex
core\cite{WangYong,Franz}. However this is in sharp contrast with
the experimental
observations\cite{Maggio,Renner,Pan,Hoogenboom,Dagan,Mitrovic},
mainly on optimally doped samples. The absence of a ZBCP within
the vortex cores was attributed to the presence of $id_{xy}$ or
$is$ components\cite{Dagan}, or the competing orders (see later).
In this paper we show that the DOS due to vortex quasiparticle
excitations deviates from Simon-Lee scaling law for the overdoped
sample, but follows rather well with the optimally doped one. The
deviations for the overdoped sample are found to be induced by the
vortex core size effect. In the extremely underdoped region, it is
found that the Simon-Lee scaling law fails, except for in very low
temperature region. This can be understood as due to the competing
order emerging within or nearby the vortex cores.

\section{Experiment}

The single crystals measured in this work were prepared by the
travelling solvent floating-zone technique. Samples with seven
different doping concentrations (p=0.063, 0.069, 0.075, 0.09,
0.11, 0.15, 0.22) have been investigated. The sample with p=0.15
and 0.22 are from CRIEP, and others are from NLSC(IOP). Part of
the data for all samples will be presented, for example the field
induced change of $\gamma$ at zero K, the residual linear term
$\gamma_0$ and the $\alpha$ value in the pure d-wave expression
$C_{DOS}=\alpha T^2$ when $H=0$ (see later). However for clarity
we mainly show data and the analysis on three typical samples with
x=0.22 ($T_c=27.4$K, overdoped), x=0.15 ($T_c=36.1$K, close to
optimal doping point) and p=0.069 ($T_c\approx 12 $K, underdoped,
x=0.063 originally) as characterized by AC susceptibility and DC
magnetization (shown by the insets in Fig.3, Fig.8 and main panel
of Fig.12). The quality of our samples has also been characterized
by x-ray diffraction patterns, and $R(T)$ data showing a narrow
transition $\Delta T_c \leq $ 2 K. For some samples, the full
width at the half maximum (FWHM) of the rocking curve of the (008)
peak is only 0.10$^\circ$. The overdoped sample has a mass about
28.55 mg and $3.66 \times 2.3 \times 0.5 mm^3$ in dimension. The
optimally doped sample weighs about 23.6 mg with dimensions of
$3.1 \times 3 \times 0.5 mm^3$. For the underdoped sample with
nominal concentration x=0.063, before annealing, it has a
superconducting transition temperature of about 12 K and a mass of
about 32.89mg and 3.75 x 2.75 x 0.5$mm^3$ in dimensions. By
fitting to the empirical relation $T_c/T_c^{max}=1-82.6(p-0.16)^2$
with $T_c^{max}=38K$ the maximum $T_c$ at the optimal doping point
$p=0.16$, we estimate that the hole concentration of this sample
is around $p=0.069$. After annealing in flowing Ar gas for 48h,
the $T_c$ drops down from about 12 K to 9 K indicating that the
sample becomes more underdoped. Note that $T_c$=9 K is expected
exactly by the empirical relation at $x=p=0.063$.

The heat capacity presented here were taken with the relaxation
method based on an Oxford cryogenic system Maglab. The sample is
put onto a micro-chip on which there is a tiny Cernox temperature
sensor, a film heater. The micro-chip together with the sample are
hung up by golden wires in vacuum. These golden wires are the only
thermal links between the micro-chip and the thermal sink whose
temperature is well controlled. The temperature of the micro-chip
is controlled by the on-board small film heater and measured by
the on-board thermometer. When the temperature of the micro-chip
is stable, a heating power with fixed current is sent to the film
heater on the chip and the time dependence of the chip temperature
is measured simultaneously. The change of temperature is fitted to
an exponential relation $\Delta T=\Delta T_0\times
[1-exp(-t/\tau)]$, and heat capacity is determined by
$\tau=(C+C_{add})/\kappa_w$, here $C$ and $C_{add}$ are the heat
capacity of the sample and addenda (including a small sapphire
substrate, small printed film heater, tiny Cernox temperature
sensor, $\phi$25 $\mu m$ gold wire leads, Wakefield thermal
conducting grease (about $100\mu g$)) respectively, where
$\kappa_w$ is the thermal conductance between the chip and the
thermal link. The value $C_{add}$ has been measured and subtracted
from the total heat capacity, thus $C$ value reported here is only
that from the sample. In Fig.1 we present the temperature and
field dependence of the heat capacity from the addenda and three
typical samples. It is clear that the heat capacity of the addenda
is much smaller than the value of the samples. In addition, the
data of $C/T$ for the addenda extrapolates to zero at $T = 0K$
showing the only existence of phonon part. One can also see that
almost no field dependence can be observed for the addenda.
However, for all samples, there is a clear finite intercept at
$T=0K$ which gives rise to a residual linear term $\gamma_0$.
Meanwhile the field induced change can be easily observed for all
samples, even for the very underdoped sample. The inter-crossing
of the data at $H=0$ and $H = 12 T$ at about 6K for the undedoped
sample is understandable and will be discussed later.

In all measurements done for the present work, the magnetic field
H is always parallel to c-axis of single crystals, and the data
are collected in the warming up process after it is cooled under a
field (Field-Cooling process, abbreviated as FC hereafter). In the
data treatment we use $\Delta \gamma=[C_{H||c}-C_{H=0}]/T$ instead
of using $\Delta \gamma=[C_{H||c}-C_{H\perp C}]/T$. The latter may
inevitably involve the unknown DOS contributions from another kind
of vortices (for example, Josephson vortices) when $H\perp C$. The
field dependence of the Cernox thermometer has been calibrated
well by Oxford before the shipment. The true temperature has been
derived automatically by the software with a calibration table
with magnetic fields at 0T, 1T, 2T, 4T, 8T and 12 T. The values at
other fields are obtained also automatically by software by doing
linear interpolation between two nearby fields. Therefore the
readout from the machine gives directly the true temperature value
with the field effect corrected.

\begin{figure}
\includegraphics[width=8cm]{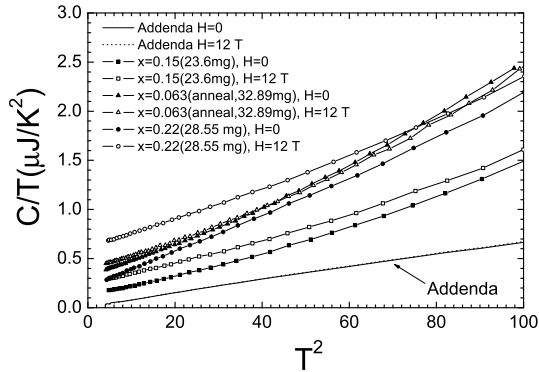}
\caption{Temperature dependence of the heat capacity from the
addenda with 110 $\mu g$ Wakefield grease (bottom solid line for
H=0 and dashed line for H=12 T), and three typical samples (filled
symbols for H=0 and open symbols for H=12 T, lines are guides to
the eye). }\label{fig1}
\end{figure}

\section{Results and Data Analysis}

\subsection{Fitting to the zero field data}

Before showing the field induced change of the heat capacity we
present in Fig.2 the temperature dependence of $C/T$ for some
samples at zero field. As mentioned previously, for a d-wave
superconductor in the superconducting state, it is known that
$C_{DOS}=\alpha T^2$ when $H=0$. In addition, as observed in other
cuprate superconductors, the curve at zero field extrapolates to a
finite value ($\gamma_0$) at 0 K instead of zero. This was
interpreted as potential scattering effect due to small amount
impurities or disorders\cite{Moler,Hirschfeld}. We will argue that
this residual linear term may also reflect physics beyond the
simple argument of impurity scattering (see later). As also
observed by other groups for $La-214$ system, the anomalous upturn
of $C/T$ due to the Schottky anomaly of free spins is very
weak\cite{Fisher,Phillips,Nohara,Chen}. This avoids the complexity
in the data analysis. Together with the phonon contribution $\beta
T^3+\delta T^5$, we have

\begin{equation}
C(H=0)/T=\gamma_0+\alpha T+ \beta T^2 + \delta T^4
\end{equation}

\begin{figure}
\includegraphics[width=8cm]{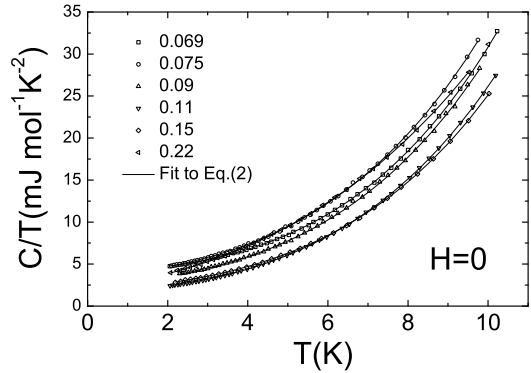}
\caption{Temperature dependence of $C/T$ for samples
(p=0.069,0.075,0.09,0.11,0.15 and 0.22) at zero field. The solid
lines are fits to eq.2 and the parameters derived here are listed
in Table-I.}\label{fig2}
\end{figure}

Above equation is used to fit the data at $H=0$ for some samples.
The fitting results are shown in Fig.2 and listed in Table-I,
where the units for $\gamma_0$, $\alpha $, $\beta $ and $\delta$
are $mJ mol^{-1}K^{-2}$, $mJ mol^{-1}K^{-3}$, $mJ mol^{-1}K^{-4}$
and $mJ mol^{-1}K^{-6}$ respectively. One can see that $\alpha$
decreases quickly towards underdoping, $\beta$ (and thus the Debye
temperature $\Theta_D$) does not change too much with doping. The
sudden drop of $\alpha$ at $p=0.11$ may be induced by the well
known $1/8$ problem. The residual linear term $\gamma_0$ increases
rapidly towards underdoping, which will be discussed later. The
$\alpha$ values are also comparable to those found by other
groups\cite{Nohara,Chen}.

\begin{table}
\caption{\label{tab:table1}}
\begin{ruledtabular}
\begin{tabular}{cccccc}
p & $T_c$ & $\gamma_0$  & $\alpha $ & $\beta $ & $\delta $ \\
\hline
0.22   & 27.4          & 2.19 & 0.463  & 0.186  & 0.00054   \\
0.15   &36.1           & 1.90 & 0.177  & 0.120  & 0.00093   \\
0.11   &29.3           & 1.70 & 0.065  & 0.137  & 0.00096   \\
0.09   &24.4           & 2.64 & 0.158  & 0.145  & 0.00110    \\
0.075  &15.6           & 3.72 & 0.131  & 0.177  & 0.00110    \\
0.069  &12.0           & 4.06 & -0.077(?) & 0.157 & 0.00117   \\

\end{tabular}
\end{ruledtabular}
\end{table}

\subsection{Overdoped sample with x=0.22}

Fig.3 shows $C/T$ as a function of $T^2$ at magnetic fields
ranging from 0 to 12 T for the overdoped sample. The separation
between each field can be well determined. In low temperature
region the curves are rather linear showing that the major part is
due to phonon contribution $C_{ph}=\beta T^3+\delta T^5$.  It is
known that the phonon part is independent on the magnetic field,
this allows to remove the phonon contribution by subtracting the
$C/T$ at a certain field with that at zero field. The results
after the substraction are shown in Fig.4. The subtracted values
$\Delta \gamma=\gamma (H)-\gamma (0) = [C(T, H)-C(T, H=0)]/T$
exhibit a rather linear $T$ dependence in low temperature region.
One can also see that the negative slope is actually field
dependent. In the following we will show that the field dependent
slope of the linear part in low temperature region shown in Fig.4
directly deviates from the Simon-Lee\cite{SimonLee} scaling law.

\begin{figure}
\includegraphics[width=8cm]{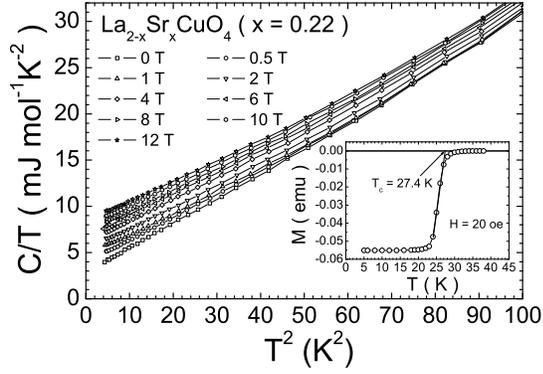}
\caption{Specific heat C/T vs. $T^2$ of the overdoped sample (x =
0.22) at magnetic fields ranging from 0 to 12 T. The inset shows
the diamagnetic transition at around 27.4 K determined by the
crossing point of the extrapolating line of the most steep part
with the normal state background M = 0.}\label{fig3}
\end{figure}

\begin{figure}
\includegraphics[width=8cm]{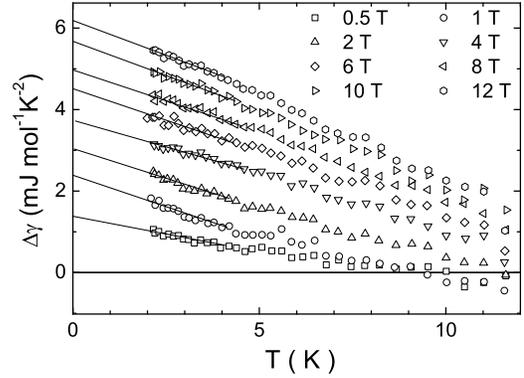}
\caption{Temperature dependence of $\Delta \gamma=
\gamma(H)-\gamma(0) = [C(H,T)-C(0,T)]/T$ of the overdoped sample
(x = 0.22). A linear behavior is clearly seen in low temperature
region with a field dependent slope, which is not in accord with
the proposed scaling law by Simon and Lee (see text). The straight
lines in low temperature region are guides to the eyes. From these
lines one can determine the zero temperature intercept
$\Delta\gamma$ and the slope $d\Delta \gamma/dT$ shown in Fig.5 }
\label{fig4}
\end{figure}

According to Simon-Lee scaling law
$C_{vol}/T\sqrt{H}=f(T/\sqrt{H})$, in low temperature region, the
Volovik's relation restores, thus one has $C_{vol}/T=A\sqrt{H}$
and further has

\begin{equation}
\Delta \gamma=[C(H)-C(0)]/T=A\sqrt{H}-\alpha T
\end{equation}

This clearly shows that there is a negative slope for $\Delta
\gamma$ vs. $T$, but the slope $\alpha$ is a constant. However,
from Fig.4 one can see that the slope changes slightly with the
magnetic field $H$. This indicates that only the Volovik's
relation is not enough to interpret the data. In the following we
will take both the core size effect and finite temperature effect
into account. The former has not been considered in the original
Simon-Lee scaling law since the size of the vortex core was
thought to be small, and the contribution from that small region
is negligible. If the vortex core size becomes bigger, this should
be reconsidered when counting the DOS due to the Doppler shift
effect.

Let us first consider only the finite temperature effect. Suppose
that we are in the crossover region between the low temperature
limit and high temperature limit suggested by Volovik and
Kopnin\cite{VolovikComment}, making Taylor's expansion to the
right hand side of Simon-Lee scaling law leads to

\begin{equation}
C_{vol}=b_0H+b_1T\sqrt{H}+b_2T^2+o(T^3)
\end{equation}

where $b_0$ = 0 because $C_{vol}/T$ should not diverge when $T=0$
and $H\neq 0$, $b1=A$. Since $o(T^3)$ is very small in low
temperature region, one thus has $C_{vol}/T=b_1\sqrt{H} + b_2 T$.
Interestingly, one can see that this simple formula contains the
results both in the low temperature limit
$C_{vol}=b_1T\sqrt{H}$\cite{Volovik,Vekhter,Kopnin1996} and high
temperature limit\cite{Vekhter,VolovikComment} $C_{vol}=b_2T^2$.
This is not surprising since a scaling function should be more
general and cover most possible cases. When $H_{c1} << H <<
H_{c2}$, the total specific heat contains four parts: Doppler
shift term from the region outside the core $C_{vol}$, the inner
vortex core term $C_{core}\propto HT$, the residual linear term
$\gamma_0T$ and the phonon term $C_{ph}$. Here it is assumed that
the heat capacity contributed by the core region is equal for each
vortex and independent on the external magnetic field, thus
$C_{core}$ depends only on the vortex density which is
proportional to H. The local DOS measured by STM\cite{Pan}
revealed that the low energy DOS within the vortex core differs
only slightly from the case for a d-wave superconductivity
(outside and far away from the vortex core). When changing the
external magnetic field, the low energy DOS within the vortex core
is not expected to vary too much. At zero field, the total
specific heat contains three parts: $\gamma_0T$ and $C_{ph}$, and
a quadratic term $\alpha T^2$ due to the thermal excitation near
the nodal region. Thus $\Delta \gamma$ can be written as:

\begin{equation}
\Delta \gamma=\gamma(H)-\gamma(0)=b_1\sqrt{H}+(b_2-\alpha
)T+b_{core}H
\end{equation}

From eq.5 one can see that $\Delta \gamma$ depends on T through
the second term, however the slope $b_2-\alpha$ is still field
independent by definition. This clearly indicates that the
Simon-Lee scaling law is still not enough to interpret the field
dependent slope of $\Delta \gamma$ vs. $T$ as shown in Fig.4.

Let us keep going, still based on eq.5, we propose that the core
size effect may have a sizable influence on the total vortex
quasiparticle excitations. This is actually reasonable since the
vortex core with size $\xi\propto\hbar v_F/\Delta_s$ grows up in
the overdoped side due to smaller superconducting gap
value\cite{Wen2003EPL}, where $v_F$ is the Fermi velocity and
$\Delta_s$ is the superconducting gap. By taking the vortex core
size ($2\xi$) into account, i.e., deducting the normal core area
away from the Volovik term, one can rewrite $\Delta \gamma$ as

\begin{equation}
\Delta \gamma=(b_1\sqrt{H}+b_2T)\times(1-\xi^2/R_a^2)-\alpha
T+b_{core}H
\end{equation}

where $\xi$ is the radius of the normal core, $R_a$ is the outer
radius of a single vortex where the supercurrent is flowing, thus
$R_a^2=\phi_0/\pi H$. Reorganizing all terms in eq.6 leads to

\begin{equation}
\Delta \gamma=b_1\sqrt{H}\times(1-\frac{\pi
\xi^2}{\phi_0}H)+(b_2-\alpha
)T-b_2\frac{\pi\xi^2}{\phi_0}HT+b_{core}H
\end{equation}

\begin{figure}
\includegraphics[width=8cm]{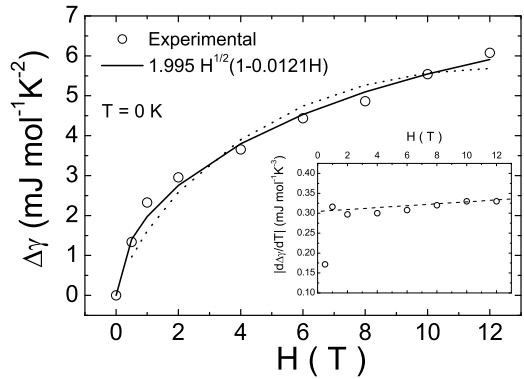}
\caption{Field induced DOS at zero K of the overdoped sample
(x=0.22). The solid line is a theoretical curve
$\Delta\gamma=1.995\sqrt{H}(1-0.0121H)$. The dotted line
represents the best fit to the case at the unitary limit
(ref.\cite{Hirschfeld} ). The inset shows the slope $d\Delta
\gamma/dT$ of the straight lines shown in Fig.4 in low temperature
region. The dashed line is a linear fit to the data at fields
above one tesla. The intercept and the slope of the dashed line
give rise to the pre-factors of the second and third terms in
eq.7} \label{fig5}
\end{figure}

One can see that the third term in eq.7 is just what we need for
interpreting the difficulty as mentioned above. It is necessary to
recall that the core size correction is proportional to $\xi^2$,
for example, it will be four times when $\xi$ doubles. Thus
increase of $\xi$ in the overdoped side will give sizable effect
on the total DOS and core size effect should be considered. Next
let us have a closer inspection at the data and derive some
parameters. At zero temperature, only the first term and the last
term are left in eq.7. The values of $\Delta \gamma(T=0)$ are
determined from the extrapolation of the linear lines in Fig.4 to
0 K and presented in Fig.5. The data $\Delta \gamma(H, T=0)$ is
also determined by doing linear fit to the raw data $C/T$ vs.
$T^2$ between $2 K$ and $4 K$, and then subtracted the zero
temperature value $\gamma_0$. The results are quite close to each
other by using these two different methods. The solid line in
Fig.5 is a fit to the data using the first term in eq.7 yielding
$b1=1.995 \pm 0.046 mJK^{-2}T^{-1/2}$ and
$\pi\xi^2/\phi_0=0.012\pm0.003$ and thus $\xi=28.2\AA$. The value
$\xi=28\AA$ derived here is quite close to that found in
Nernst\cite{WangYaYu} and STM measurements\cite{Pan} ($20\AA$ for
optimally doped Bi-2212 sample). We also tried to use the first
term together with the last term to fit the data but find out that
the contribution from the last term is extremely small. {\em The
first term here describes the zero temperature data very well,
indicating the absence of a second component of order parameter
such as $id_{xy}$ or $is$ since otherwise the Fermi surface would
be fully gapped and the Doppler shift had very weak effect on the
quasiparticle excitations}. The inset of Fig.5 shows the field
dependence of the slope of the linear part in Fig.4. It is clear
that the slope increases roughly linearly with H above 1 T. This
can be exactly anticipated by the second and third terms in eq.7.
From the inset of Fig.5 one obtains $\alpha-b_2=0.305mJ mol^{-1}
K^{-3}$ and $b_2\pi\xi^2/\phi_0=0.00238 mJ mol^{-1}K^{-3}T^{-1}$.
By taking $\xi = 28.2 \AA$, we obtain the following values:
$\alpha=0.501 mJ mol^{-1}K^{-3}$ and $b_2=0.196 mJ
mol^{-1}K^{-3}$. The value of $\alpha=0.501 mJ mol^{-1}K^{-3}$
found here is quite close to the value obtained by fitting the
zero field data to eq.2 (0.465 $mJ/mol K^2$ see Table-I). Since
the contribution from the core region (last term in eq.7) is
negligible comparing to the Volovik term, from eq.7 one
understands that the failure of using the Simon-Lee scaling law in
very overdoped sample is due to the core size effect. This is
actually quite reasonable since the core size in the overdoped
region grows up. Sofar we don't know yet whether the negligible
contribution from the core region is because of the gapped feature
within the core region as found in optimally and underdoped
samples\cite{Maggio,Renner,Pan,Hoogenboom,Dagan,Mitrovic}, or it
is naturally small comparing to the contributions from the Doppler
shift effect of the surrounding superfluid. This casts a
interesting issue for future STM measurement on the tunnelling
spectrum within the vortex cores in very overdoped region.

It is necessary to estimate how much of the field induced
delocalized DOS is contributed by the impurity scattering in our
present sample. At the unitary limit at zero energy, i.e., when
T=0, K$\ddot{u}$bert and Hirschfeld\cite{Hirschfeld} predict that
the field induced relative DOS is
$\delta\gamma/\gamma_0=P_1(H/P_2)log(P_2/H)$, where
$P_1=0.322(\Delta_0/\Gamma)^{1/2}$, $\Delta_0 $ the gap maximum,
$\Gamma$ the impurity scattering rate, and $P_2=\pi H_{c2}/2a^2$,
$a\approx 1$. The dotted line in Fig.5 represents the best fit of
this relation to our data yielding $\Gamma/\Delta_0\approx0.00039$
(close to the clean limit). In addition, the value of $H_{c2}$
derived here is about 21.53T, which is too small for present
sample. It is clear that the fit has a poor quality comparing to
the better fit in the clean limit (solid line). Furthermore, the
formulism considering the impurity effect does not predict a field
dependent slope for the linear relation $\Delta\gamma$ vs. T in
low temperature region. Therefore together with the extremely
small $\Gamma/\Delta_0$ found in present case, we believe that the
field induced DOS in our sample comes mainly from the Doppler
shift effect on supercurrent outside the cores. The residual
linear term ( $\gamma_0T$ ) of electronic specific heat will be
discussed separately in the following subsection.

\begin{figure}
\includegraphics[width=8cm]{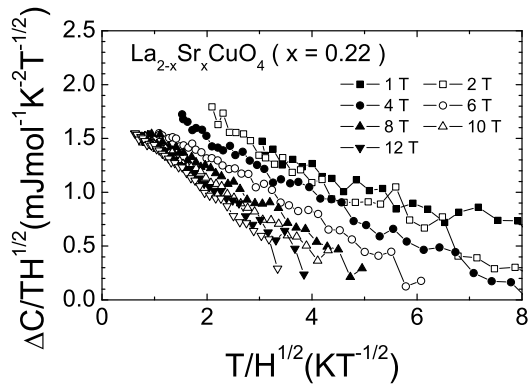}
\caption{ Scaling of the raw data $[C(H)-C(0)]/T\sqrt{H}$ vs.
$T/\sqrt{H}$ for the overdoped sample based on the Simon-Lee
scaling. Clearly no good scaling can be obtained.} \label{fig6}
\end{figure}

In order to show the inapplicability of Simon-Lee scaling law for
the overdoped sample, we present the raw data
$[C(H)-C(0)]/T\sqrt{H}$ vs. $T/\sqrt{H}$ in Fig.6. Clearly the
scaling looks very poor. From above discussion, we conclude that
the failure of the Simon-Lee scaling law in the very overdoped
region is due to the quite large vortex core size which needs to
be corrected.

\begin{figure}
\includegraphics[width=8cm]{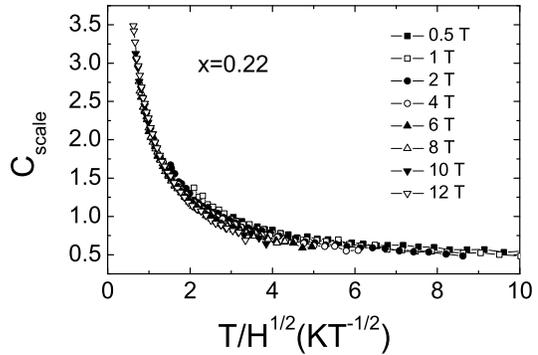}
\caption{ Plot of the data $C_{scal}=(\Delta
C/T^2+\alpha)/(1-\pi\xi^2H/\phi_0)$ vs. $T/\sqrt{H}$. It is clear
that the data collapse onto one main branch which is expected by
the theoretical expression with core size correction (eq.8).}
\label{fig7}
\end{figure}

The nice fit in Fig.5 with only the first term of eq.7 suggests
that the core region has very small contribution to the DOS since
otherwise the last term $b_{core}H$ should be sizeable. This
implies that the low energy DOS inside the vortex core is very
small. Based on this idea, we write a new scaling law as

\begin{equation}
\frac{\Delta C/T^2+\alpha}{1-\pi\xi^2H/ \phi_0 } =
\frac{\sqrt{H}}{T}f(\frac{T}{\sqrt{H}})
\end{equation}

One can use this equation to test the idea about the vortex core
size correction. We thus present the data of $(\Delta C/T^2
+\alpha)/(1-\pi\xi^2H/\Phi_0)$ vs. $T/\sqrt{H}$ in Fig.7 with
$\alpha =0.501 mJ /mol K^3$ and $\pi\xi^2/\phi_0=0.0121$ derived
above. The data collapse on one branch and show good consistency
with the expected theoretical curve. The slight scattering or
deviation from the main scaling branch is due to the simple
assumption made for the core size correction ($1-\xi^2/R_a^2$) and
the rough estimation for $\alpha$ value. Worthy of noting is that
to have this nice data collapsing and consistency with the
theoretical curve we need to take $b_{core}\approx 0$, again
showing a small contribution from the inner vortex core.  The nice
data collapsing using eq.8 suggests that the Simon-Lee scaling law
can be reconciled by considering the vortex core size effect. It
is interesting to note that the electronic thermal conductivity
derived by Sun et al.\cite{SunXF} is not consistent with the
Volovik's expression in low temperature region for the overdoped
sample, rather it shows a plateau when the field is high. However
the $H^{1/2}$ law is followed very well in low temperature region
for the optimally doped sample. Our core-size correction picture
may provide alternative interpretation to this discrepancy.

\subsection{Optimally doped sample (x=0.15)}

\begin{figure}
\includegraphics[width=8cm]{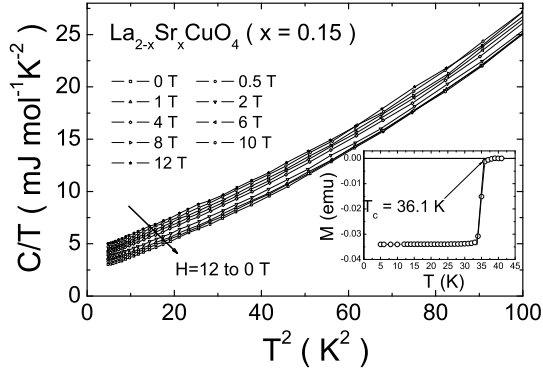}
\caption{ Raw data of $C(H)/T$ vs. $T^2$ for the optimally doped
sample. The inset shows the diamagnetic transition measured in the
ZFC mode at $20 Oe$.} \label{fig8}
\end{figure}

\begin{figure}
\includegraphics[width=8cm]{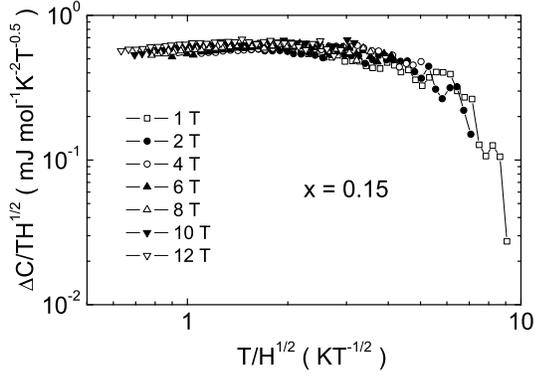}
\caption{ Scaling of the raw data $[C(H)-C(0)]/T\sqrt{H}$ vs.
$T/\sqrt{H}$ for the optimally doped sample (x = 0.15). The
scaling looks rather good.} \label{fig9}
\end{figure}

\begin{figure}
\includegraphics[width=8cm]{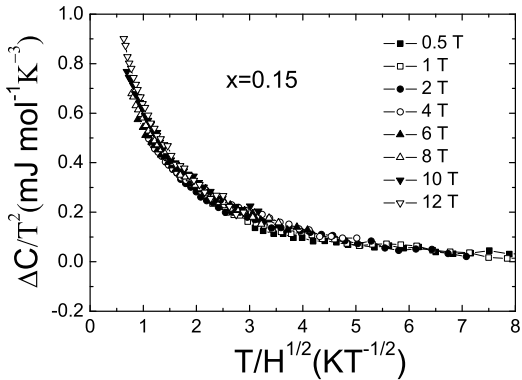}
\caption{Scaling of the raw data $[C(H)-C(0)]/T^2$ vs.
$T/\sqrt{H}$ for the optimally doped sample based on the Simon-Lee
scaling law. Clearly the scaling quality is quite good.}
\label{fig10}
\end{figure}

\begin{figure}
\includegraphics[width=8cm]{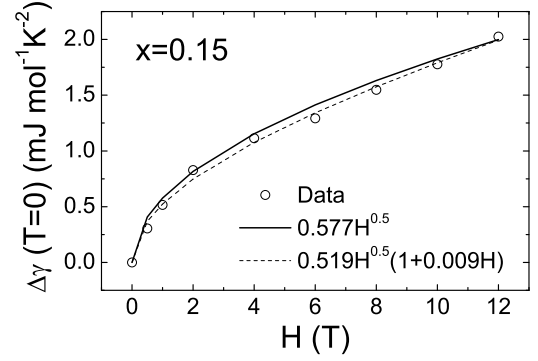}
\caption{Zero temperature specific heat $\Delta \gamma(H)$ of the
optimally doped sample. The solid line is a theoretical expression
$\Delta\gamma(H)= 0.577\sqrt{H}$ $(mJ/mol K^2)$ which fits the
data very well. The dashed line is a fit to the first term of eq.7
yielding a small and unreasonable negative value for
$\pi\xi^2/\phi_0$.} \label{fig11}
\end{figure}

In order to have a comparison with the overdoped sample, in this
subsection we present the data from an optimally doped one with
x=0.15. The raw data of specific heat for the optimally doped
sample is shown in Fig.8. The separation between each field can
also be easily distinguished in low temperature region. Again here
the curve at zero field extrapolates to a finite value
($\gamma_0$) at 0 K instead of zero. This will be discussed in the
forthcoming subsection. It is found that the linear behavior of
$\Delta\gamma$ vs. $T$ for the overdoped sample (shown in Fig.4)
is absent here. This may be due to the much smaller $\alpha$ value
(see Table-I). We then check whether the d-wave scaling law is
applicable here. If the Volovik (Doppler shift) effect really
dominates here, one can expect that $C(H)-C(0)=
T^2y(T/\sqrt{H})-\alpha T^2$, thus $[C(H)-C(0)]/T^2$, or
$[C(H)-C(0)]/T\sqrt{H}$ should scale with $T/\sqrt{H}$. In Fig.9
we present the result of $[C(H)-C(0)]/T\sqrt{H}$ vs. $T/\sqrt{H}$.
It is clear the scaling is rather good comparing to that of the
overdoped sample [Fig.6]. Here the value of $\Delta
\gamma/\sqrt{H}$ in the zero temperature limit gives the
pre-factor $A$ in the Volovik's relation $C_{vol}=AT\sqrt{H}$
which is about 0.55 to 0.6 $mJ mol^{-1}K^{-2}T^{-1/2}$. In Fig.10
we present the Simon-Lee scaling in the way of $[C(H)-C(0)]/T^2$
vs. $T/\sqrt{H}$. One can see that the scaling is reasonably good.
All data below about 10 K collapse onto one branch. We have been
aware that Nohara et al.\cite{Nohara} successfully used the
Simon-Lee scaling law to the overdoped sample $x=0.19$, but failed
for the optimally doped one. The failure of using Simon-Lee
scaling law in Nohara's experiment for optimally doped sample is
in contradiction with the reports from many other
groups\cite{Chen,Fisher,Phillips}. This may be caused by the way
that they used to derive $\Delta \gamma$. As stressed in previous
subsection, we use $\Delta \gamma=[C_{H||c}-C_{H=0}]/T$ instead of
using $\Delta \gamma=[C_{H||c}-C_{H\perp C}]/T$ to derive the
field induced change of $\gamma$. The latter (as used by Nohara et
al.) may inevitably involve the unknown DOS contributions from
another kind of vortices (for example, Josephson vortices) when
$H\perp C$. For $La-214$ system, since the Schottky anomaly is
very week, it is not necessary to derive $\Delta \gamma$ in the
second way. While Nohara et al.\cite{Nohara} obtained a relatively
good scaling for the overdoped sample ($x=0.19$). We would not
comment on the validity of this successful scaling at $x=0.19$.
One reason for the discrepancy between their results and our
results may be from different doping levels: our sample ($x=0.22$)
is more overdoped and the vortex core size is certainly larger and
needs to be corrected .

The data $\Delta \gamma(H, T=0)$ is determined by doing linear fit
to the raw data $C/T$ vs. $T^2$ between $2 K$ and $4 K$, and then
subtracted the zero temperature value $\gamma_0$. The results are
shown in Fig.11. We tried to fit the zero temperature data in
Fig.11 to the first term in eq.7 in terms of core size correction
( shown by the solid line), it turns out that the correction term
$\pi\xi^2H/\phi_0$ is very small and negative, which is certainly
unreasonable. This actually indicates that $\Delta \gamma(T=0)$
can be nicely fitted to the theoretical expression $\Delta
\gamma(H)= 0.577\sqrt{H} (mJ/mol K^2)$. Using
$C_{vol}/T=k\gamma_n\sqrt{H/H_{c2}}$, we have
$k\gamma_n=0.577\sqrt{H_{c2}}$. A similar value (0.49) was derived
by Fisher et al.\cite{Fisher} for $La-214$ sample with $x=0.15$.
Taking $H_{c2} \approx 100 T$\cite{Wen2003EPL} and $k \approx$
0.74\cite{Hussey}, we have $\gamma_n\approx 7.8 mJ/mol K^2$, which
is very close to the reported values for optimally doped $La-214$
sample\cite{Nohara,Chen}. Worthy of noting here is that the field
induced extra DOS at zero K can be nicely fitted with the
Volovik's relation $C_{vol}\propto T\sqrt{H}$ albeit the residual
linear term $\gamma_0$ is quite large. This suggests that the
residual linear term $\gamma_0$ observed commonly in curate
superconductors may originate from some other properties, such as
inhomogeneity. It may not be induced by the small scale impurity
scattering since otherwise the $H^{1/2}$ law should not be
followed so well.  Since both the Simon-Lee scaling law and the
Volovik's $\sqrt{H}$ are followed very well for optimally doped
sample, the core size effect seems to be very week.

\subsection{Underdoped sample}

In this subsection the low temperature specific heat of underdoped
$La_{2-x}Sr_xCuO_4$ ($p=0.069, 0.075, 0.09$ and $0.11$) single
crystals is reported in magnetic fields up to 12 T. It is found
that the Volovik's relation $C_{vol}=ATH^{1/2}$ is still satisfied
in the zero temperature limit, but the proposed Simon-Lee scaling
law, i.e., $C_{vol}/T^2=f(T/\sqrt{H})$, is not followed so well,
except for at very low temperatures (below about 3-4 K).

Fig.12 shows the temperature dependence of the AC susceptibility
 and DC magnetization of the underdoped sample $La_{2-x}Sr_xCuO_4$ ($p=0.069$). The
transition temperature drops from about 12 K to 9 K after
extracting some oxygen out of the sample (not shown here) by
annealing the sample in $Ar$ gas for 48 hrs. Then $T_c$ keeps
stable upon further annealing in $Ar$ gas. The $T_c$ is increased
again when the sample is treated in flowing oxygen. The DC
magnetization measured in the FC process shows a transition width
of about $2.5$K. Below about $7.5 K$ the M(T) curve keeps flat.
The magnetization measured in ZFC mode shows a slight increase
with temperature induced by the easy flux penetration in the very
underdoped region. Specific heat has been measured in the FC mode
as done for all other samples. This mode provides a vortex system
which is close to equilibrium state and thus relatively
uniform\cite{LiSL}. Presented in Fig.13 is the specific heat $C/T$
as a function of $T^2$ at magnetic fields ranging from 0 to 12T
for this underdoped sample before annealing (estimated p=0.069).
In low temperature region the curves are rather linear showing
that the major part is due to the phonon contribution
$C_{ph}=\beta T^3+\delta T^5$, and has no slight upturn in low
temperature region due to the Schottky anomaly of free spins. The
curve at zero field extrapolates to a finite value ($\gamma_0$) at
0K , again showing the existence of a residual linear term which
will be discussed later. As mentioned before the phonon part is
independent on the magnetic field, this allows to remove the
phonon contribution by subtracting the C/T at a certain field with
that at zero field, one has $\Delta C=C(H)-C(0)=C_{vol}-\alpha
T^2$ and $\Delta C/T^2=C_{vol}/T^2-\alpha$. The results after the
substraction are shown in Fig.14. One can see that the linear part
with negative slope as appearing for the overdoped sample is
absent here. This is understandable when $\alpha$ value (or
$\alpha T^2$ term) is very small comparing to the field induced
change of total specific heat. Therefore for this underdoped
sample, no apparent $T^2$ term at $H=0$ was observed which can be
found easily in the overdoped LSCO sample. This is consistent with
the data shown in Table-I and experimental results from other
groups on LSCO\cite{Chen,Nohara}. The disappearance of this
$\alpha T^2$ term was usually interpreted as due to either the
impurity scattering which smears out the nodal effect, or the
small value of coefficient $\alpha$ of the $T^2$ term. We will
show that this is induced by much smaller $\gamma_n$ value (thus
smaller $\alpha$) just above $T_c$.

\begin{figure}
\includegraphics[width=8cm]{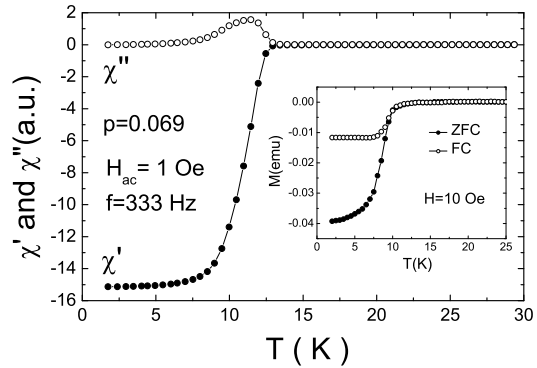}
\caption{ AC susceptibility and DC magnetization of the underdoped
sample with p=0.069. The bottom curve is the real part $x'$ of the
ac susceptibility, and the upper one is the imaginary part $x''$.
By extrapolating the most steep transition portion of the real
part of the AC susceptibility to the normal state background
($x'=0$), the $T_c=12 K$ is determined here. The inset shows the
DC magnetization measured in the FC and ZFC processes. Below about
7.5 K, the magnetization measured in FC mode is rather stable. The
slight temperature dependence of the magnetization measured in ZFC
mode is induced by the flux penetration. } \label{fig12}
\end{figure}

\begin{figure}
\includegraphics[width=8cm]{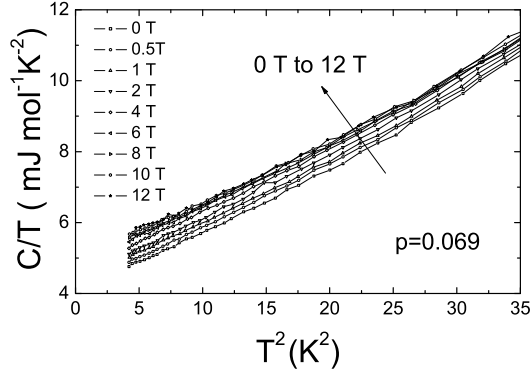}
\caption{Raw data of $C(H)/T$ vs. $T^2$ for the underdoped sample
p=0.069. One can see that the field induced change of specific
heat becomes much smaller than that of the optimally and overdoped
sample.} \label{Fig13}
\end{figure}

\begin{figure}
\includegraphics[width=8cm]{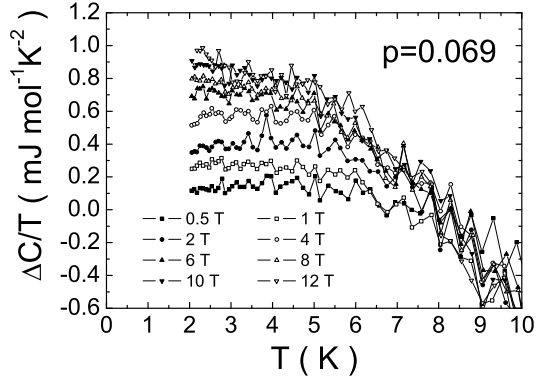}
\caption{The subtracted data [C(H)-C(0)]/T  vs. T for the
underdoped sample ($p=0.069$). It is clear that no linear part
with negative slope of $\Delta\gamma$ vs. T as appearing for the
overdoped sample can be observed here. This may be induced by the
much smaller $\alpha$ value. } \label{Fig14}
\end{figure}

\begin{figure}
\includegraphics[width=8cm]{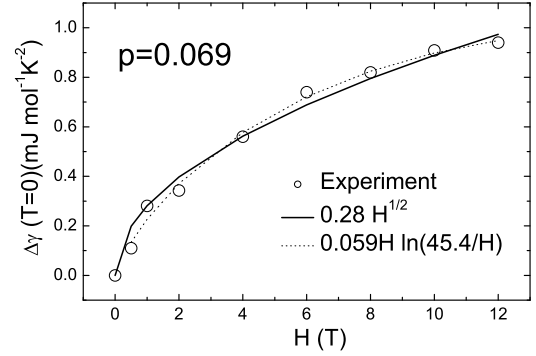}
\caption{Field induced change of $\gamma$ at zero K for the
underdoped sample with p=0.069. The solid line and dashed line are
fits to the theoretical relation of Volovik effect at the clean
limit and the impurity scattering at the unitary limit
respectively.} \label{Fig15}
\end{figure}

Next let us have a look at the field induced DOS at T=0 K. The
data $\Delta \gamma(H, T=0)$ is determined by doing linear fit to
the raw data $C/T$ vs. $T^2$ between $2 K$ and $4 K$, and then
subtracted the zero temperature value $\gamma_0$. The results are
shown in Fig.15. In order to compare with the theoretical
predictions, the increase in $\gamma(H, T=0)$ was fitted with
$\Delta \gamma(H, T=0)= AH^B$, and the value of $B$ is $0.52$ and
$A$ is about 0.28. The value $B$ derived here from free fitting is
very close to $0.5$ as predicted by the Volovik theory
\cite{Volovik} which may manifest the existence of line node in
the gap function. We can also fix $B=0.5$ and find out that
$A=0.282 mJ/mol K^2T^{1/2}$. This is also compatible with the
results of other groups\cite{Chen}. For the zero temperature data
we also considered the core size correction, i.e., tried to use
the first term of eq.7 to fit the zero temperature data. But it
gives rise to a small and negative value of $\pi\xi^2/\phi_0$
which is certainly unreasonable.

\begin{figure}
\includegraphics[width=8cm]{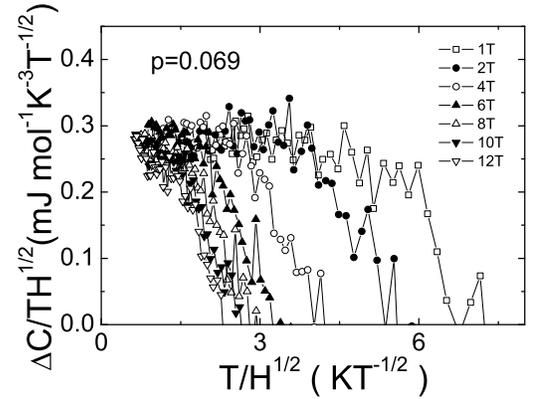}
\caption{Scaling of the raw data $\Delta C/T\sqrt{H}$ vs.
$T/\sqrt{H}$ for the sample $p=0.069$ based on the Simon-Lee
scaling. Clearly no good scaling can be obtained, except for at
very low temperatures.} \label{fig16}
\end{figure}

\begin{figure}
\includegraphics[width=8cm]{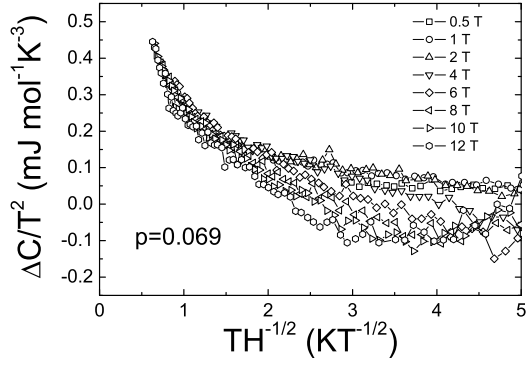}
\caption{Scaling of the raw data $\Delta C/T^2=[C(H)-C(0)]/T^2$
vs. $T/\sqrt{H}$ for the sample $p=0.069$ based on the Simon-Lee
scaling. Good scaling can be found only at very small values of
$T/\sqrt{H}$. } \label{fig17}
\end{figure}

For the underdoped sample, we used the Simon-Lee scaling law to
scale our data. The results of $[C(H)-C(0)]/T\sqrt{H}$
vs.$T/\sqrt{H}$ are shown in Fig.16. The data fan out showing a
poor scaling quality. Clearly the data cannot be scaled using the
Simon-Lee scaling law except for at very low temperatures. We plot
also the data of $[C(H)-C(0)]/T^2=C_{DOS}/T^2-\alpha$ vs.
$T/\sqrt{H}$ in Fig.17, one can again see the poor scaling in wide
temperature region. The Simon-Lee scaling has been applied to all
samples investigated in this work (p= 0.069, 0.075, 0.09, 0.11,
0.15, 0.22). It is easy to find that the scaling quality becomes
better and better when the doping concentration is increased from
0.069 to 0.15. One can even see the gradual change among these
underdoped samples (p=0.069, 0.075, 0.09, 0.11): the scaling
curves fan out like that in Fig.17 for samples with $p=0.069,
0.075$, but the scaling pattern becomes narrower towards higher
doping. The scaling behavior are shown in Figs. 18-20 for samples
with $p=0.075$, $p=0.09$ and $p=0.11$. A clear trend for a better
scaling at a higher doping can be easily seen here.

\begin{figure}
\includegraphics[width=8cm]{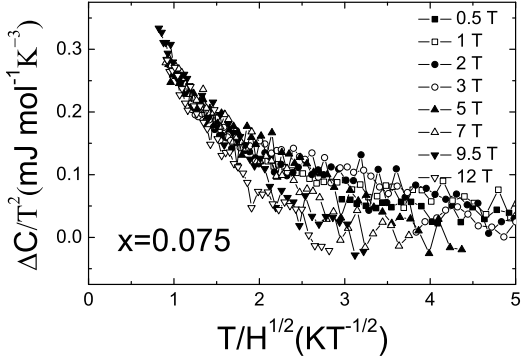}
\caption{Scaling of the raw data $\Delta C/T^2=[C(H)-C(0)]/T^2$
vs. $T/\sqrt{H}$ for the sample $p=0.075$ based on the Simon-Lee
scaling. Good scaling can be found only at very small values of
$T/\sqrt{H}$. } \label{fig18}
\end{figure}

\begin{figure}
\includegraphics[width=8cm]{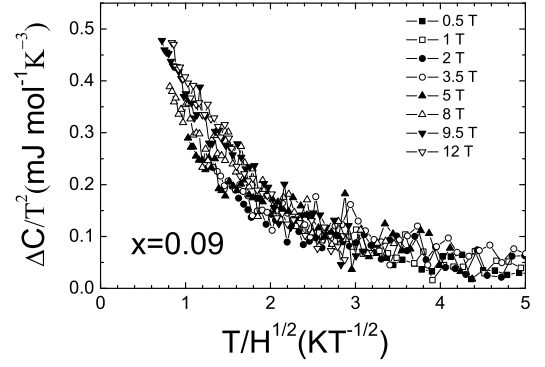}
\caption{Scaling of the raw data $\Delta C/T^2=[C(H)-C(0)]/T^2$
vs. $T/\sqrt{H}$ for the sample $p=0.09$ based on the Simon-Lee
scaling. } \label{fig19}
\end{figure}

\begin{figure}
\includegraphics[width=8cm]{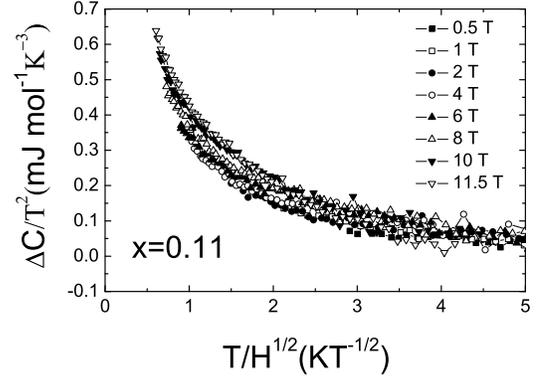}
\caption{Scaling of the raw data $\Delta C/T^2=[C(H)-C(0)]/T^2$
vs. $T/\sqrt{H}$ for the sample $p=0.11$ based on the Simon-Lee
scaling. Now the fanning-out of the scaling curves are strongly
constrained showing a better scaling behavior. } \label{fig20}
\end{figure}

There are several possibilities for the failure of using Simon-Lee
scaling law in underdoped region. One possibility is due to the
impurity scattering effect as suggested by K$\ddot{u}$bert et
al.\cite{Hirschfeld}. Thus we use the dirty limit formula
$\gamma(H)=\gamma(0)[1+D(H/H_{c2})ln(H_{c2}/H)]$ to fit the data
at zero K, where $D\approx\Delta/32\Gamma$. For the simplicity we
show here only the fit to the data of the sample $p=0.069$. It is
found that the data can also be roughly fitted by the relation
with impurity scattering (as shown by the dashed line in Fig.15).
The obtained results for the sample with $p=0.069$ are:
$H_{c2}=45.6 T$, $\gamma(0)=4.03$mJ/mol $K^2$,
$\Gamma/\Delta=0.046$. Thus it seems that one cannot rule out the
possibility of impurity scattering to play a dominant role in the
field induced change of $\gamma$ in very underdoped sample.
However this speculation cannot interpret the nice $\sqrt{H}$
dependence of the field induced DOS at zero K as shown in Fig.15.
Worthy of noting is that the dirty limit formula of
K$\ddot{u}$bert et al.\cite{Hirschfeld} is more flexible to fit to
the data than the simple $\sqrt{H}$ relation. One needs to seek an
alternative way to clarify this discrepancy.

The second possibility is the core size effect as appearing in the
overdoped sample. We then try to scale the data by using eq.8 and
leaving both $\alpha$ and $\pi\xi^2/\phi_0$ as free fitting
parameters. Unfortunately no good scaling can be found by choosing
any values for $\alpha$ and $\pi\xi^2/\phi_0$. This is in
consistent with the fact that an unreasonable negative value for
$\pi\xi^2/\phi_0$ is obtained if we fit the zero temperature data
in Fig.15 to the first term of eq.7. Both indicate that the
failure of Simon-Lee scaling law here is not due to the core size
effect. One may argue that the data is scalable with only a very
narrow scaling region of $T/\sqrt{H}$, for example, from
Fig.16-18, the scalable region is about $T/\sqrt{H} \le 1.5
KT^{-0.5}$. This is of course possible since we don't know the
precise value for many parameters. However we can have a rough
estimation to check whether this is reasonable. Provided the
scalable region is $T/\sqrt{H} \le T_c/\sqrt{H_{c2}(0)}=1.5$,
inputting $T_c=12 K$, one has $H_{c2}(0)$=64 T which seems too big
for this very underdoped sample.

Another possible reason for the failure of the scaling law is that
the sample is in the underdoped region with a pseudogap in the
normal state. When the sample is in the mixed state, some
competing or coexisting order such as short range
antiferromagnetic
order\cite{Lake,Kang,ZhangSC,Arovas,ZhuJX,ChenY,Franz2}, or the
SDW order\cite{Sachdev}, or a d-density wave (DDW) order\cite{DDW}
is enhanced, and this newly generated or enhanced order will
certainly give contribution to the total specific heat. For
example, for 2D AF correlation, it is known that $C_{AF}\propto
T^2$. Therefore qualitatively the failure of the Simon-Lee scaling
law in underdoped region can be understood in the following way.
By increasing the magnetic field, a second order is generated or
enhanced within the vortex core and nearby regions (about
100$\AA$). On one hand this region is gapped leading to the
decrease of the total DOS at fermi level simply by reducing the
region where the supercurrent can flow. On the other hand the
newly generated AF or SDW or DDW region will contribute a new term
to the total specific heat due to spin or other type excitation.
The relevant competing order under a magnetic field, according to
both neutron scattering\cite{Lake,Kang} and NMR
measurement\cite{NMR}, may be the AF order. STM measurement by
Hoffman et al.\cite{Hoffman} indicates a checkerboard like
modulation with periodicity of $4a$ of the LDOS. This was regarded
as the direct observation of the strong electronic correlation
with the underlying competing order which was predicted by many
theoretical
work\cite{Arovas,ZhuJX,ChenY,Franz2,Zaanen,Low,Sachdev2,ZhangSC2}.
This qualitative picture calls for further detailed analysis and
evidence from other experiments. Since the heat capacity from the
newly generated or enhanced second order has a temperature
dependence of $T^{\epsilon}$ with $\epsilon > 1$, at zero
temperature the specific heat from this term is zero, thus the
$H^{1/2}$ law from the Doppler shift of the d-wave
superconductivity is restored. This may be the reason for that the
zero-temperature data follows the $H^{1/2}$ law but the data at
finite temperatures do not satisfy the Simon-Lee scaling very
well.

\subsection{The residual linear term $\gamma_0$}

\begin{figure}
\includegraphics[width=8cm]{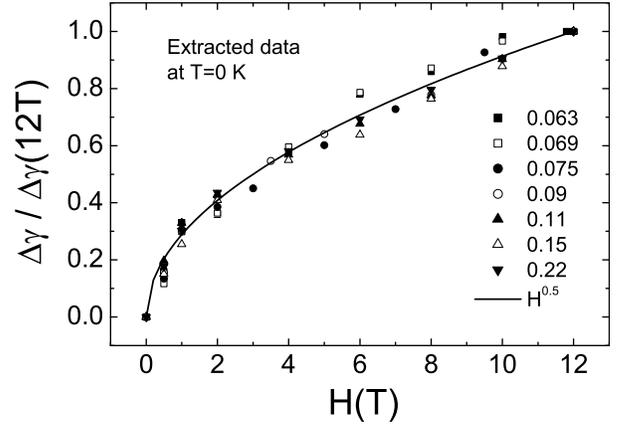}
\caption{Field dependence of the field induced extra $\gamma$
normalized by the value at 12 T at zero K. The solid line
represents the theoretical curve $H^{1/2}$. It is found that the
data from different samples at different doping levels follow the
$H^{1/2}$ law reasonably well}.\label{fig21}
\end{figure}

\begin{figure}
\includegraphics[width=8cm]{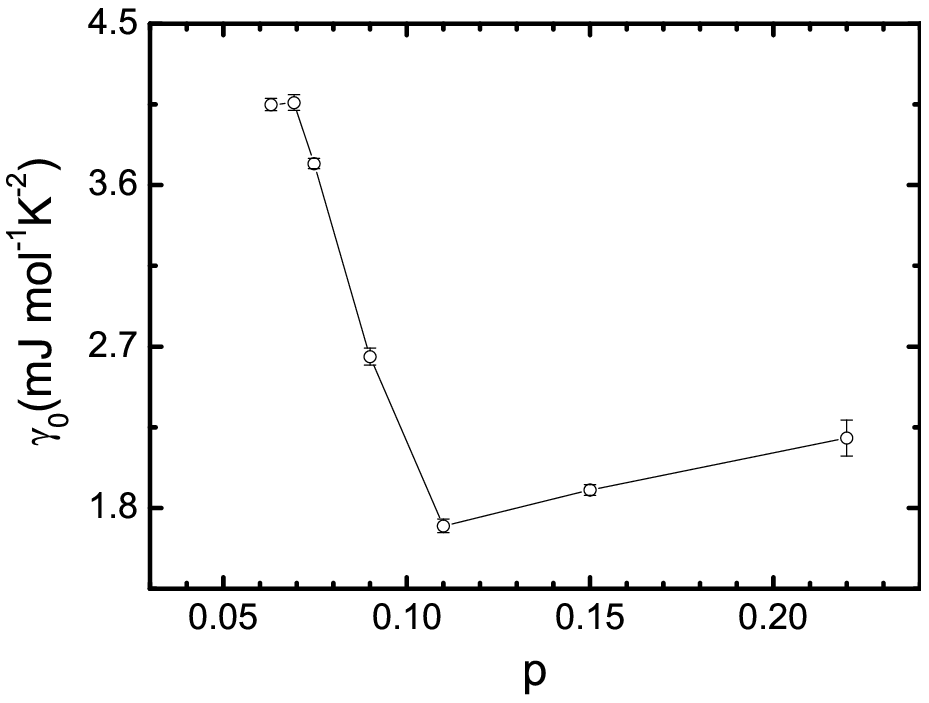}
\caption{The doping dependence of $\gamma_0$ of
$La_{2-x}Sr_xCuO_4$} crystals. It is clear that the minimum
$\gamma_0$ value is found in the region around p =0.11 to 0.125
.\label{fig22}
\end{figure}

Almost in all cuprate superconductors, a residual linear term of
electronic specific heat $\gamma_0$ has been observed in the low
temperature limit $T\rightarrow$0, even in the best samples up to
date. In $YBa_2Cu_3O_7$ single crystals, Moler et al.\cite{Moler}
found that $\gamma_0$ is larger for the twinned samples than the
detwinned ones. Meanwhile they further found that $\gamma_0$
increases when the sample becomes more underdoped. Clearly one can
conclude that $\gamma_0$ is related to the impurities or disorders
in the samples. While quite surprisingly, for many samples with
quite different $\gamma_0$ values, it is found that the zero
temperature data can be expressed as
$\gamma(H,T=0)=\gamma_0+A\sqrt{H}$, showing evidence for d-wave
pairing. This may suggest that $\gamma_0$ is not mainly induced by
the impurity scattering, since otherwise the field induced extra
DOS should not follow the relation $\Delta C(H, T=0)/T=A\sqrt{H}$
so well. In Fig.21, we present the field dependence of $\Delta
\gamma=[C(H,T=0)-C(0, T=0)]/T$ normalized to the value for each
sample at about 12 T. Meanwhile we show the $H^{1/2}$ law by the
solid line. One can see that for almost all samples, the field
induced extra DOS at zero K follows the $H^{1/2}$ relation
reasonably well despite the $\gamma_0$ value highly disperses.
This feature was also discovered by Chen et al.\cite{Chen} on
three typical samples ($x=0.10, 0.16$ and $0.22$). Nohara et
al.\cite{Nohara} measured three single crystals (x = 0.10, 0.16
and 0.19) and empirically found that the optimally doped sample (x
= 0.16) has the lowest value of $\gamma_0$ ($\gamma_0$=2.8, 1.5
and 2.2 $mJ mol^{-1}K^{-2}$ for x=0.10, 0.16 and 0.19 respectively
obtained from Fig.1 of Ref.\cite{Nohara}). Chen et al.\cite{Chen}
found the similar behavior among three samples with $x=0.10, 0.16
$ and $ 0.22$ ($\gamma_0$=1.49, 0.7 and 1.41 $mJ mol^{-1}K^{-2}$
respectively). This raises the question of the origin of this
residual linear term and its correlation with the field induced
quasiparticle DOS. As mentioned before, if the field induced DOS
is related to the impurity scattering, another
relation\cite{Hirschfeld}
$\delta\gamma/\gamma_0=P_1(H/P_2)log(P_2/H)$ is expected. This is
some time contradicting to the experimental result (see dotted
line in the main panel of Fig.5). In addition, the $\sqrt{H}$
dependence of the field induced change of $\gamma$ is certainly
not obtained by accident, since it is found on different samples
from different groups, even on poly-crystalline
samples\cite{Chen}. From the point of view of chemistry it is not
true that the optimally doped sample is the cleanest one since in
most cases the underdoped samples can be more easily grown with
high quality. In this sense the residual linear term may be
related to some other properties rather than the impurity
scattering. In Fig.22 the $\gamma_0$ values for different single
crystals measured in our experiment from underdoped to overdoped
are shown. The value of $\gamma_0$ is obtained by fitting zero
field data to eq.2 (see Table-I). It is clear that the minimum
$\gamma_0$ is found in the region around 0.11 or 0.125. The value
for $\gamma_0$ found from our data are more close to the data of
Nohara et al.\cite{Nohara} on single crystals, but clearly higher
than that obtained on polycrystalline smples\cite{Chen}. So far we
don't know the reason for this discrepancy. For our extremely
underdoped sample (x = 0.063) investigated here, although the data
at finite temperatures cannot be treated with the Simon-Lee
scaling law, the data in the low temperature limit $T \rightarrow
0$ can be however still nicely expressed by
$\gamma(H,T=0)=\gamma_0+A\sqrt{H}$, even the absolute increase of
$\gamma(H,T=0)$ by field is much smaller than $\gamma_0$.
Therefore it is reasonable to conclude that the field induced part
is mainly contributed by the Doppler shift effect on the
supercurrent outside the vortex cores, while the residual linear
term $\gamma_0$ is mainly contributed by some small normal regions
which weakly depends on the magnetic field. Similar explanations
to the origin of $\gamma_0$ have been suggested for many times in
the past\cite{Phillips2}. This may be understood in the following
way. In underdoped $Bi-2212$ single crystals,
scanning-tunneling-microscopic (STM) measurement indeed reveal a
mixture of superconducting regions with sharp quasiparticle
coherent peaks on the tunnelling spectrum, and the
non-superconducting regions with pseudogap-like tunneling
spectrum\cite{STM}. In the overdoped side, the tiny normal cores
as proposed in the Swiss cheese model\cite{Uemura2}, or the
mesosscopic normal regions suggested by Fukuzumi et
al.\cite{Fukuzumi} and Wen et al.\cite{Wen} will contribute a
residual term $\gamma_0$ which does not show an apparent increase
with the field. As proposed by Fukuzumi et al.\cite{Fukuzumi} that
the dome-like electronic phase diagram may be formed by the
mixture of three phases: anti-ferromagnetic phase in the extremely
underdoped region, a d-wave superconducting region with the robust
superconductivity near the optimal doping point and a
non-superconducting Fermi liquid in the overdoped region.
According to this simple picture the $\gamma_0$ should increase in
the underdoped and overdoped region, which is just the case as
shown by the data in Fig.22. Therefore we would argue that the
residual linear term may be mainly contributed by some
non-superconducting regions due to phase separation, either
chemical or electronic in origin. This interesting argument needs
certainly to be further checked with data obtained by different
techniques on different systems.

\subsection{Field induced reduction of specific heat in high temperature region}

In above analysis, we concentrate on the data below 10 K ( below 6
K for the very underdoped sample). This is also the temperature
region that most of the low temperature specific heat data was
reported in the literatures. Now we report another phenomenon:
field induced reduction of specific heat in the mixed state. In
Fig.23 we present the temperature dependence of the field (12T)
induced change of $\gamma$ for three typical samples analyzed
above, here $\Delta \gamma=[C(12T)-C(0T)]/T$. Although the data
are strongly scattered one can still see that: (1) The field
induced change $\Delta \gamma$ becomes negative at about $0.5-0.7
T_c$; (2) The curves have a similar shape: $\Delta\gamma$ is
positive in low temperature region, then it becomes negative and
finally comes back to zero in high temperature region (near $T_c$
for optimal and overdoped sample). For the overdoped sample, the
$\Delta \gamma$ keeps negative above $0.5T_c$ until $T_c$ at which
$\Delta \gamma$ suddenly goes back to zero. For the optimally
doped one, the $\Delta\gamma$ is negative above about $0.7T_c$ up
to the highest temperature we measured here (30 K). However for
the underdoped sample, it shows that the $\Delta \gamma$ keeps
negative until $1.5 T_c$. Similar data were obtained by Fisher et
al.\cite{Fisher} on samples with $x=0.15$. Our data near $T_c$ is
more scattered since our setup can only measure samples with
maximum mass of 50 mg. This feature, namely the negative $\Delta
\gamma$ in high temperature region is a consequence of entropy
consideration, which has been observed in all types of
superconductors. In low temperature region, when a magnetic field
is applied, vortices will be generated leading to higher DOS near
Fermi surface, so that $\Delta \gamma=\gamma(H)-\gamma(0)$ is
positive. When the temperature is increased, to satisfy the field
independent entropy above $T_c$, in a certain region below $T_c$,
$\Delta \gamma$ should be negative. The most interesting point is
for the underdoped sample here, even above $T_c$ one clearly sees
a magnetic field induced change of entropy. This implies an
abnormal normal state which is far from a conventional metal. For
a conventional s-wave superconductor, the field induced change of
$\gamma$ can be negative near $T_c$. It is difficult to understand
the field induced reduction of specific heat well below $T_c$,
since the normal core region always gives rise to a higher DOS of
quasiparticle. Outside the vortex core the DOS is almost
negligible. However this field induced reduction of specific heat
well below $T_c$ is found to be a general feature of all LSCO
crystals we investigated so far. This may be related to the
intrinsic properties of cuprate superconductors. In a d-wave
superconductor, theoretically it is predicted that there is a ZBCP
within the vortex core which should also contribute a quite high
DOS\cite{WangYong,Franz} in the mixed state. Besides, a high DOS
will be generated by the Doppler shift effect of the supercurrent
surrounding the vortex core. Normally the sum of these two terms
are larger than the zero-field term $C_{DOS}=\alpha T^2$, leading
to a field induced enhancement of DOS in low temperature region.
When the temperature is high, the Doppler shift effect will be
smeared out by the strong thermal excitation and finally $\Delta
\gamma$ becomes negative. As far as we know, no quantitative
theoretical expression about $\Delta \gamma$ has been reported so
far for a d-wave superconductor in wide temperature region. We
cannot have a quantitative understanding to our data. However,
this field induced reduction of specific heat well below $T_c$ may
be understood as due to the anomalous feature of vortex core
state, i.e., a gapped vortex core as seen by the STM.\cite{Pan},
or based on the assumption that the contributions from the core
region is much smaller than the outside region where either the
Doppler shift or the strong thermal excitation dominates. Actually
the Simon-Lee scaling law becomes a $C_{vol}\propto T^2$ relation
in high temperature region. In this case the quasiparticle
excitation outside the vortex core is almost the same ($\alpha
T^2$) with or without applying a magnetic field. However since the
vortex core region is gapped or contributes negligible part to the
total DOS, one needs to take the core region away from the total
area in calculating $\Delta \gamma$, naturally leading to a
negative value of $\Delta \gamma$.

\begin{figure}
\includegraphics[width=8cm]{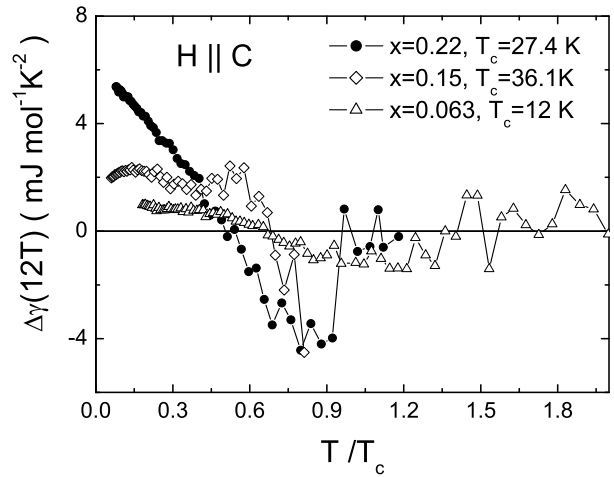}
\caption{Temperature dependence of the field (12T) induced change
of $\Delta\gamma$ = $[C(12T)-C(0T)]/T$ vs. T for three typical
samples (x=0.22, overdoped: filled circles; x=0.15, optimally
doped: diamonds; x=0.069, underdoped: triangles ). The horizontal
axis is normalized to $T_c$ of each sample. In high temperature
regime below $T_c$, the field induced change of DOS becomes
negative.}\label{fig23}
\end{figure}

\section{Discussion}
In low temperature region, our analysis indicates that the field
induced quasiparticle DOS can be well described by Volovik's
theory or Simon-Lee scaling law although a correction due to the
core size effect is needed for the overdoped sample. This means
that the prerequisite for the theory, i.e., the $d_{x^2-y^2}$
pairing symmetry is well satisfied. Therefore it naturally rules
out the presence of a second order parameter like $id_{xy}$ or
$is$ either due to overdoping\cite{YehNC} or due to the field
effect\cite{Laughlin} in all samples investigated here. Meanwhile,
for the overdoped sample, another interesting phenomenon is that
the vortex core region contributes very little (at least much
smaller than that induced by the Doppler shift if the
super-current would flow in the same area) to the total DOS. We
have also tried to analyze the data of the optimally doped and
underdoped sample in the way as that for overdoped one, for
example to fit the data in Fig.11 and Fig.15 to the first term in
eq.7. It turns out however that the correction term
$\pi\xi^2H/\Phi_0$ derived is small and negative which is
unreasonable. For the optimally doped sample, it is quite easy to
understand since the vortex core becomes very small. However for
the underdoped sample, it is quite hard to understand since the
core size tends to grow up too\cite{Wen2003EPL}. The negligible
contribution from the vortex core region may suggest that the ZBCP
is absent within the cores, even in the overdoped region. This
suggestion inferred from the specific heat measurement about the
ZBCP within the vortex core is consistent with the tunnelling
results\cite{Maggio,Renner,Pan,Hoogenboom,Dagan} and certainly
clears up the concerns about the surface conditions in the STM
measurement. Recent results from NMR also show the absence of a
ZBCP inside the vortex core\cite{Mitrovic}. In this sense our data
together with the earlier NMR data present a bulk evidence for an
anomalous vortex core. Interestingly it is widely perceived that
the normal state in overdoped region shows a Fermi liquid behavior
even when the superconductivity is completely
suppressed\cite{Proust}. If this is the case the mean-field frame
of BdG theory based on the conventional d-wave superconductivity
seems not enough to interpret the anomalous vortex core state in
HTS. For the underdoped sample, the Simon-Lee scaling fails except
for in very low temperature region. This is interpreted as due to
the presence of a second (gapped) order like AF or SDW or DDW
within and nearby the vortex core. However one needs more
theoretical and experimental efforts to show the justice of this
argument.

By fitting the field induced extra DOS at zero temperature to the
relation $\Delta \gamma =AH^{1/2}$, we obtained the pre-factor $A$
in wide doping regime, where $A=0.74
\gamma_n/\sqrt{H_{c2}}$\cite{Hussey}. The results are presented in
Fig.24. It is seen that the A-value increases with the doping
concentration monotonously. This can be understood in the
following way: by increasing doping the normal state value
$\gamma_n$ will increase\cite{Loram}, the $H_{c2}$ will drop down
(at least it is the case in the overdoped region). Therefore
A-value will increase monotonously in the overdoped side. One can
see from the data that the A-value keeps almost constant in the
extremely underdoped region, which means that $\gamma_n$ and
$H_{c2}$ should both decrease with underdoping. This indicates
that the $H_{c2}$ becomes smaller and the coherence length $\xi$
becomes larger towards more underdoping. This is consistent with
the recent conclusion drawn by Wen et al.\cite{Wen2003EPL} by
analyzing the data about the low temperature flux dynamics. This
conclusion about the coherence length calls for a direct check to
the vortex core size by using scanning-tunnelling-microscopy in
the future.

\begin{figure}
\includegraphics[width=8cm]{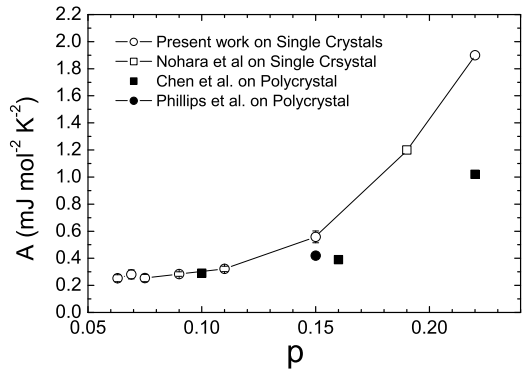}
\caption{The doping dependence of the pre-factor A in
$\gamma(T=0)=\gamma_0+AH^B, B\approx 0.5$. It is evident that the
A-value increases with the hole concentration monotonously. The
data measured on polycrystalline samples are somewhat smaller,
which is perhaps induced by the random orientation of the grains.
For some grains the field is not parallel to the $c-axis$ leading
to a smaller contribution to the field induced change of $\gamma$.
}\label{fig25}
\end{figure}

\section{Concluding Remarks}

In conclusion, the field induced change of the electronic specific
in mixed state of a series $La_{2-x}Sr_xCuO_4$ single crystals has
been measured and extensively analyzed. It is found that the field
induced DOS of the optimally doped sample fits the predicted
Simon-Lee scaling law for a d-wave superconductor very well, while
deviations have been found for the overdoped sample. However, it
is reconciled for the overdoped sample if one considers the core
size effect provided the contribution from the inner vortex core
is small comparing to that due to the Doppler shift in the same
area. The Simon-Lee scaling law is applicable in the underdoped
region only in very low temperature region. We attribute this to
the appearance of a second competing order (like AF or SDW or DDW)
within and nearby the vortex core. The negligible contribution
from the vortex core region may suggest the absence of the ZBCP in
the vortex core, even in the overdoped region, although it is
expected by the Bogoliubov de-Gennes theory for a d-wave
superconductor. Finally we present the doping dependence of the
residual linear term $\gamma_0$ commonly observed in cuprate
superconductors. It is argued that this linear term may be related
to inhomogeneity (either electronic or chemical), rather than be
simply explained as due to the small scale impurity scattering as
usually thought. This conclusion is made because the field induced
extra DOS at zero temperature follows the Volovik's $\sqrt{H}$ law
reasonably well in all doping regime. It is hard to believe that
this nice consistency is obtained by accident. Our results
generally conclude a d-wave pairing symmetry for the hole doped
$La_{2-x}Sr_xCuO_4$ samples, although some competing orders may
co-exist with the superconductivity, and an anomalous feature
(missing of the ZBCP) may appear within the vortex core.

\section{Acknowledgments}

This work is supported by the National Science Foundation of China
, the Ministry of Science and Technology of China, the Knowledge
Innovation Project of Chinese Academy of Sciences. We thank
Xiaogang Wen, Dunghai Lee, Ashvin Vishwanath, Subir Sachdev, Jan
Zaanen, C. S. Ting, Pengcheng Dai, Zhengyu Weng, Qianghua Wang,
and Zidan Wang for fruitful discussions.


Correspondence should be addressed to hhwen@aphy.iphy.ac.cn

\end{document}